\documentclass[a4paper,fleqn,usenatbib]{mnras}

\usepackage{newtxtext,newtxmath}

\usepackage[T1]{fontenc}
\usepackage{ae,aecompl}

\usepackage{graphicx}	
\usepackage{amsmath}	

\usepackage{multirow}
\usepackage{bm}
\usepackage[dvipsnames]{xcolor}

\newcommand{\dd}{\mathrm{d}}

\newcommand*{\be}{\begin{equation}}
\newcommand*{\ee}{\end{equation}}
\newcommand{\bea}{\begin{eqnarray}}
\newcommand{\eea}{\end{eqnarray}}

\usepackage{soul}

\title[Constraining the neutrino mass]{Constraining the neutrino mass using 
a multi-tracer combination of
two galaxy surveys and CMB lensing}

\author[M.~Ballardini, R.~Maartens]{Mario Ballardini$^{1,2,3,4}$\thanks{Contact e-mail: \href{mailto:mario.ballardini@inaf.it}{mario.ballardini@inaf.it}}, 
Roy Maartens$^{4,5,6}$\\
$^{1}$Dipartimento di Fisica e Astronomia, Alma Mater Studiorum Universit\`a di Bologna, via Gobetti 93/2, I-40129 Bologna, Italy\\
$^{2}$INAF/OAS Bologna, via Piero Gobetti 101, I-40129 Bologna, Italy\\
$^{3}$INFN, Sezione di Bologna, via Irnerio 46, I-40126 Bologna, Italy\\
$^{4}$Department of Physics \& Astronomy, University of the Western Cape, Cape Town 7535, South Africa \\
$^{5}$Institute of Cosmology \& Gravitation, University of Portsmouth, Portsmouth PO1 3FX, UK\\
$^6$National Institute of Theoretical \& Computatiopnal Sciences (NITheCS), South Africa}

\pubyear{2021}

\begin{document}
\maketitle

\begin{abstract}
Measuring the total neutrino mass is one of the most exciting opportunities available with next-generation 
cosmological data sets. We study the possibility of detecting the total neutrino mass  using large-scale 
clustering in 21cm intensity mapping and photometric galaxy surveys, together with CMB information.
We include the scale-dependent halo bias contribution due to the presence of massive neutrinos, and use a 
multi-tracer analysis in order to reduce cosmic variance.
The multi-tracer combination of an SKAO-MID 21cm intensity map with Stage~4 CMB dramatically 
shrinks the uncertainty on total neutrino mass to $\sigma(M_\nu) \simeq 45\,$meV, using only linear clustering 
information ($k_{\rm max} = 0.1\, h/$Mpc) and without a prior on optical depth. When we add to the multi-tracer 
the clustering information expected from LSST, the forecast is $\sigma(M_\nu) \simeq 12\,$meV.
\end{abstract}




\section{Introduction} \label{sec:intro}

Massive neutrinos leave unique imprints on cosmological observables throughout the history of 
the Universe \citep[see][for reviews]{Hannestad:2006zg,Lesgourgues:2006nd,Wong:2011ip,Lesgourgues:2012uu,Lattanzi:2017ubx}.
As a result, cosmology is likely to deliver  the first experimental measurement of the total neutrino mass $M_\nu \equiv \sum m_\nu$.

Cosmological data from {\em Planck} 2018 \citep{Akrami:2018vks,Aghanim:2018oex} 
in combination with BOSS DR12 clustering information \citep{Alam:2016hwk}, 
currently provide the constraint 
 $M_\nu < 120\,$meV at 95\% confidence level (CL) \citep{Aghanim:2018eyx,Vagnozzi:2017ovm,Ivanov:2019hqk}.
The current best neutrino mass limit is $M_\nu < 90\,$meV (95\% CL) \citep{DiValentino:2021hoh}, which
uses also the additional information from Pantheon Type Ia supernovae and eBOSS DR16 cosmological measurements.

These limits are model-dependent and usually weaken in cosmologies 
beyond the $\Lambda$CDM+$M_\nu$ model, in particular for cosmologies with 
extended dark energy models or modified theories of gravity 
\citep{Vagnozzi:2017ovm,RoyChoudhury:2019hls,Ballardini:2020iws,Sekiguchi:2020igz}.

On the other hand, future cosmic microwave background (CMB) anisotropy observations from ground-based 
experiments and satellites, such as
CMB-S4\footnote{\href{https://cmb-s4.org/}{https://cmb-s4.org}}, 
LiteBIRD\footnote{\href{http://litebird.jp/eng/}{http://litebird.jp}}and the 
Simons Observatory\footnote{\href{https://simonsobservatory.org/}{https://simonsobservatory.org}}, 
together with large-scale structure surveys that will be performed by 
DESI\footnote{\href{https://www.desi.lbl.gov}{https://www.desi.lbl.gov}}, 
Euclid\footnote{\href{https://www.euclid-ec.org}{https://www.euclid-ec.org}}, 
Roman Space Telescope\footnote{\href{https://wfirst.gsfc.nasa.gov/index.html}{https://wfirst.gsfc.nasa.gov}}, 
Rubin Observatory\footnote{\href{https://www.lsst.org}{https://www.lsst.org}}, 
SKA Observatory\footnote{\href{https://www.skatelescope.org}{https://www.skatelescope.org}}, 
and others, promise a robust detection of the cosmological neutrino mass also for extended models 
\citep{Allison:2015qca,Villaescusa-Navarro:2015cca,Schmittfull:2017ffw,Sprenger:2018tdb,Brinckmann:2018owf,Yu:2018tem,Boyle:2018rva}.

Neutrinos can travel cosmological distances during structure formation, modifying  
halo formation on large scales
and inducing scale-dependence of the halo bias  
around the neutrino free-streaming scale. This changes the relation between the halo number 
density and the matter (CDM+baryon) density contrast.
The effect was predicted by 
\citet{Villaescusa-Navarro:2013pva,Castorina:2013wga,LoVerde:2014pxa}, and measured 
in N-body simulations by \cite{Villaescusa-Navarro:2013pva} and \cite{Chiang:2017vuk,Chiang:2018laa}.

\citet{LoVerde:2016ahu} proposed a very promising way to target the detection of the 
cosmological neutrino mass using the so-called {\em multi-tracer} approach \citep{Seljak:2008xr}
and focusing on this scale-dependent feature in the halo bias.
Halo bias is particularly interesting because it is a quantity that is not subject 
to cosmic variance when combining the information coming from different tracers of large-scale structure.

While the importance of the scale-dependent bias due to massive neutrinos and its 
consequences for parameter inference has been investigated and stressed in \cite{Raccanelli:2017kht} 
and \cite{Vagnozzi:2018pwo}, in our work we investigate the possibility 
to improve the neutrino mass measurement through the  combination and cross-correlation 
of radio, optical, and microwave cosmological observations.

This paper is organized as follows: in \autoref{sec:theory} we briefly review the  scale-dependent features 
imprinted on the halo bias in cosmologies with massive neutrinos. We summarize the key specifications for 
the radio (SKAO-MID Band~1 and the futuristic PUMA 21cm intensity mapping), optical (LSST), and microwave 
(CMB-S4) surveys considered in our analysis in \autoref{sec:methodology}. 
We discuss the cross-correlation coefficient between CMB lensing and large-scale structure clustering 
in \autoref{sec:cross}.
Finally, we present our results in \autoref{sec:results} and we draw our conclusions in \autoref{sec:conc}.

\section{Scale-dependent bias from massive neutrinos} \label{sec:theory}
On very large scales, there is a linear relationship between fluctuations in the number density 
of halos, $\delta_h \equiv \delta n_h/n_h$, and fluctuations in the underlying density field, 
$\delta_X \equiv \delta \rho_X/\rho_X$ \citep{LoVerde:2014pxa}:
\begin{equation}
    \delta_h = b\delta_X\,,
\end{equation}
where $X$ denotes baryons ($b$), cold dark matter (CDM, $c$), massive neutrino ($\nu$), or a combination of them.

On scales above the baryonic Jeans scale, baryons and CDM behave 
indistinguishably and can be treated as a single fluid, with energy density 
$\rho_{bc} \equiv \rho_b + \rho_c$, and number density contrast
\begin{equation}
    \delta_{bc} = 
    \frac{\Omega_b \delta_b + \Omega_c \delta_c}{\Omega_b + \Omega_c} \,.
\end{equation}
The halo bias defined with respect to the CDM\,+\,baryon fluid, 
\begin{equation}
    b_{bc} = \frac{\delta_h}{\delta_{bc}}\,,
\end{equation}
is scale independent on large scales and universal, since halo formation is governed 
by local processes only \citep{Kaiser:1984sw,Bardeen:1985tr,10.1093/mnras/262.4.1065,Mann:1997df}.
Scale-dependent features might arise due to properties inherited in galaxy formation and evolution, 
and they naturally appear on small scales \citep{Castorina:2013wga}.

\begin{figure}
\centering\includegraphics[width=0.9\linewidth]{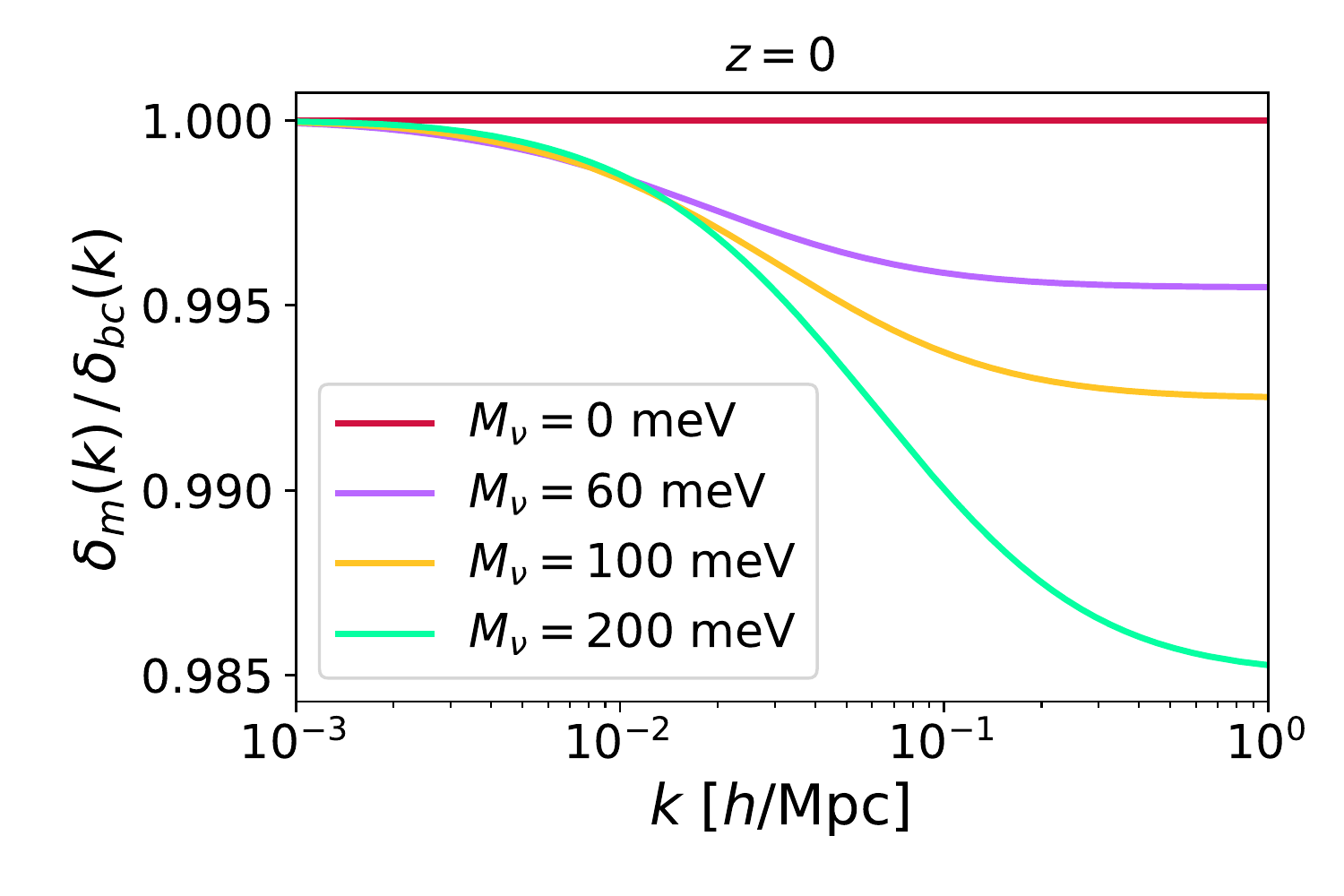}
\centering\includegraphics[width=0.9\linewidth]{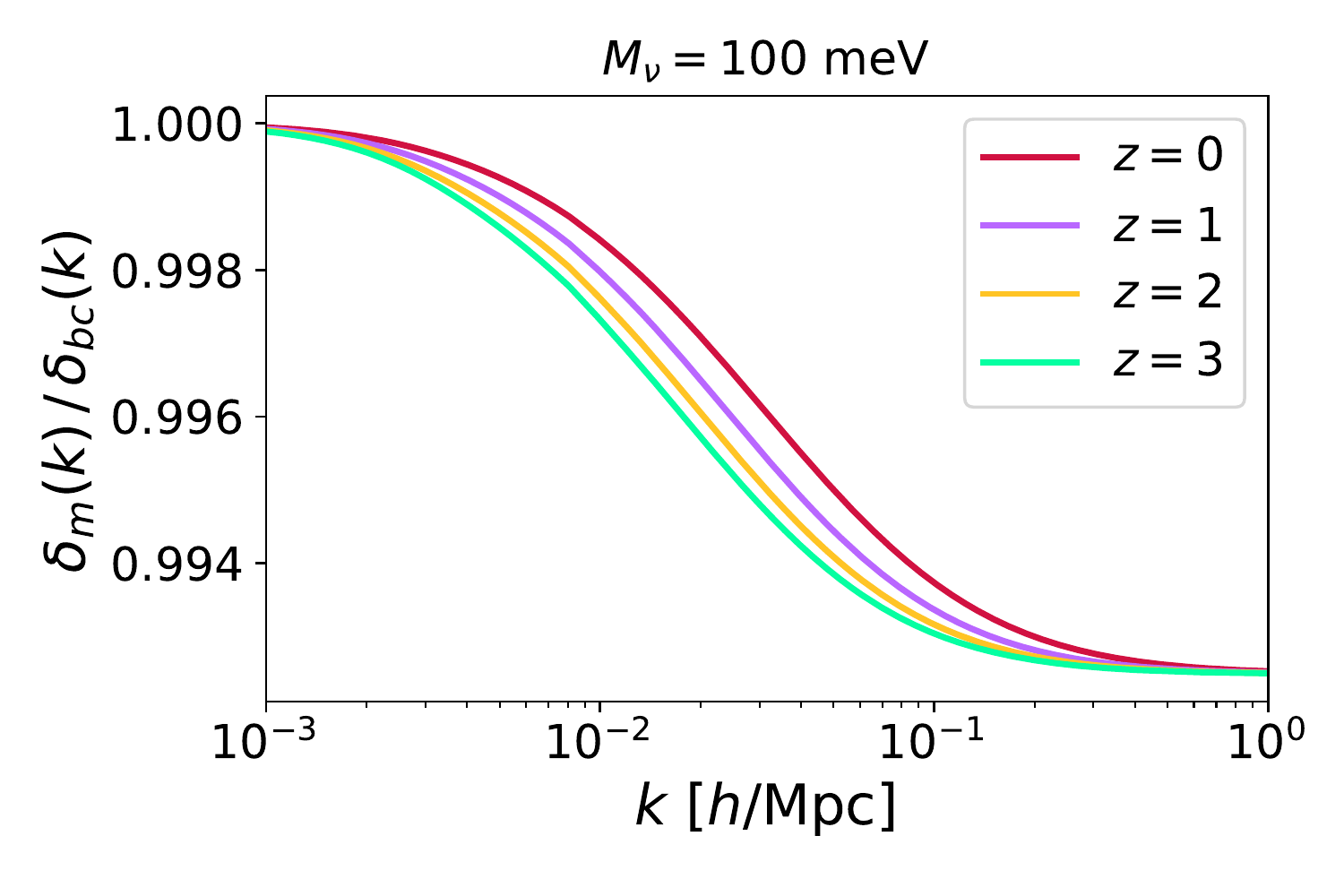}
\caption{Scale-dependence induced in  the matter
density contrast in the presence of a single massive neutrino.
{\em Top:} At $z=0$,
with  $M_\nu=0,\,60,\,100,\,200$ meV. 
{\em Bottom:} With $M_\nu=100$ meV, at $z=0,\,1,\,2,\,3$.}
\label{fig:ratio_transfer}
\end{figure}

If we define the halo bias relative to the total matter field
\begin{equation}
    b_m = \frac{\delta_h}{\delta_m}\,,
\end{equation}
where
\begin{equation}
    \delta_m = 
    \frac{\Omega_{bc} \delta_{bc} + \Omega_\nu \delta_\nu}{\Omega_m} = 
    (1-f_\nu)\delta_{bc}+f_\nu\delta_\nu\,,
\end{equation}
then this manifests a scale-dependent feature from massive neutrinos, with a step-like 
behaviour, around the neutrino free-streaming scale 
\citep{Eisenstein:1997jh,Hu:1997vi,Castorina:2013wga,Villaescusa-Navarro:2013pva,LoVerde:2014pxa}. 
This is illustrated in \autoref{fig:ratio_transfer}.

On scales larger than the neutrino free-streaming scale, and when neutrinos are non-relativistic, the {\em bc} and $\nu$ fluids 
are tightly coupled, $\delta_{bc} \approx \delta_\nu$, leading to $\delta_m \approx \delta_{bc}$. 
On smaller scales, neutrino perturbations are damped and neutrinos do not cluster, 
so that $\delta_m \approx (1-f_\nu)\delta_{bc}$.
The amplitude of the feature is larger in cosmologies with larger neutrino masses 
(see \autoref{fig:ratio_transfer}), and for more massive halos 
\citep{LoVerde:2014pxa,Raccanelli:2017kht}.

\section{Methodology and simulated datesets} \label{sec:methodology}
{Following \citet{Ballardini:2019wxj}}, we use the Fisher matrix formalism to 
forecast constraints on the cosmological parameters, assuming that the observed fields are Gaussian 
random distributed.

The Fisher matrix at the power spectrum level is then
\begin{equation} \label{eqn:fisher}
    F_{\alpha\beta} = f_{\rm sky} \sum_{\ell=\ell_{\min}}^{\ell_{\rm max}} \left(\frac{2\ell+1}{2}\right)\,
    \text{tr} \big[{\bm C}_{\ell,\alpha}\,\bm{\Gamma}^{-1}_\ell \, {\bm C}_{\ell,\beta}\,\bm{\Gamma}^{-1}_\ell\big]\,,
\end{equation}
where ${\bm C}_{\ell}=[C_\ell(z_i,z_j)]$ is the covariance  matrix, 
${\bm C}_{\ell,\alpha}=\partial {\bm C}_{\ell}/\partial \theta_\alpha$ is its derivative 
with respect to the   cosmological parameter $\theta_\alpha$, and 
$\bm{\Gamma}_{\ell}={\bm C}_{\ell}+\bm{\mathcal N}_{\ell}$ is the 
observed covariance, with $\bm{\mathcal N}_{\ell}$ the diagonal noise matrix.
This equation assumes that all experiments observe the same patch of sky, 
{with the same $f_{\rm sky}$}. 

We consider for each experiment its own sky fraction, while for the cross-correlations we use the 
estimated overlapping sky fractions. In particular, 
{we assume} $f_{\rm sky} = 0.4$ for CMB-S4, 
$f_{\rm sky} \simeq 0.48$ for SKAO-MID, 
$f_{\rm sky} = 0.5$ for PUMA, 
$f_{\rm sky} \simeq 0.33$ for LSST, and 
$f_{\rm sky} = 0.4$ for CMB-S4 $\times$ SKAO-MID, 
$f_{\rm sky} = 0.4$ for CMB-S4 $\times$ PUMA, 
$f_{\rm sky} \simeq 0.33$ for CMB-S4 $\times$ LSST, 
$f_{\rm sky} \simeq 0.33$ for SKAO-MID $\times$ LSST, 
$f_{\rm sky} \simeq 0.33$ for PUMA $\times$ LSST.
We calculate the Fisher matrices over the common patch and then we add them together.

The angular power spectra are 
\begin{equation} \label{eqn:APS}
    C_{\ell}^{XY}(z_i,\,z_j) = 4 \pi \int \frac{\dd k}{k}\, {\cal P_R} (k)\, I_\ell^{X} (k,\,z_i) \, I_\ell^{Y} (k,\,z_j) \,. 
\end{equation}
Here $X,Y = {\rm T}, {\rm E}, \phi$ for the CMB, and $=\Delta_g$ or $\Delta_{\rm HI}$ for the 
galaxy or intensity mapping (IM) surveys of post-reionisation  neutral hydrogen (HI),  where $\Delta_g=\delta_g+\,$observational corrections from observing on the 
past lightcone, and similarly for $\Delta_{\rm HI}$ \citep[see][for details]{Challinor:2011bk,Alonso:2015uua,Alonso:2015sfa,Fonseca:2015laa,Ballardini:2018cho,Ballardini:2019wxj}. 
${\cal P_R}$ is the dimensionless primordial power spectrum and the large-scale structure 
kernels are
\begin{align}
    I_\ell^{\Delta_g} (k,\,z_i) &= \int \dd z\, n^i_g(z)\, \Delta^g_\ell(k,\,z) \,, \\
    I_\ell^{\Delta_{\rm HI}} (k,\,z_i) &= \int \dd z\  W_{\rm th}(z,\,z_i)\, \bar{T}_{\rm HI}(z)\, \Delta^{\rm HI}_\ell(k,\,z)\,,   
\end{align}
where $\Delta^g_\ell, \Delta^{\rm HI}_\ell$ are the angular transfer functions 
\citep{Ballardini:2018cho}, and $W_{\rm th}(z,\,z_i)$ is a smoothed top-hat window function for 
the $i$-th bin 
{to ensure numerical stability. 
$n_g(z_i)$} includes a Gaussian window over the average angular number density $\bar{n}_g$ (see below), and $\bar{T}_{\rm HI}$ is the average brightness temperature.
We refer the reader to \citet{Hu:1997hp} for the details of the CMB temperature and polarization 
window functions.

The standard cosmological parameter vector that we use is 
\begin{equation}\label{parvec}
    \text{\boldmath$\theta$} = \left\{\omega_b,\omega_c,H_0,\tau,\ln \big(10^{10}A_s\big),n_s,M_\nu\right\} .
\end{equation}
We also include a pair of nuisance parameters for each redshift bin, in each of the large-scale structure surveys, 
allowing for a free redshift evolution of the clustering bias $b_g$, or of the combination 
${\bar{T}_{\rm HI} b_{\rm HI}}$ for IM, and for a free redshift evolution of the galaxy magnification bias $s_g$.

The fiducial cosmology used for the standard cosmological parameters follows 
{\em Planck} 2018 \citep{Aghanim:2018eyx}: 
$\omega_b = 0.022383$, 
$\omega_c = 0.12011$, 
$H_0 = 67.32$ km/s/Mpc, 
$\tau = 0.0543$, 
$\ln \big(10^{10}A_s\big) = 3.0448$, 
$n_s = 0.96605$, 
${M}_\nu = 60$ meV.
We assume one massive and two massless neutrinos with $N_{\rm eff} = 2.046$.
All angular power spectra are calculated using a modified version of the publicly 
available code\footnote{\href{https://github.com/cmbant/CAMB}{https://github.com/cmbant/CAMB}}  {\tt CAMB}
\citep{Lewis:1999bs,Howlett:2012mh,Challinor:2011bk}.
Small-scale non-linear corrections to the matter power spectrum are modelled with the {\tt Halofit} 
model \citep{Bird:2011rb,Takahashi:2012em}.

\subsection{Radio survey: single-dish mode}
We consider intensity maps of the  21cm emission of neutral hydrogen. For the fiducial linear bias model 
and background  HI brightness temperature,  we use the fitting formulas \citep{Santos:2017qgq}: 
\begin{eqnarray}
    b_{\rm HI}(z)&=&
0.667 + 0.178\, z + 0.0502\,  z^2\,,  \\ 
    \bar{T}_{{\rm HI}}(z) 
&=& 0.0559 + 0.232\, z - 0.0241\, z^2 ~   \text{mK}.
\end{eqnarray}

The noise variance for IM with $N_{\rm dish}$ dishes in single-dish mode in the frequency
$i$-channel, assuming scale-independence and no correlation between the noise in different frequency 
channels, is \citep{Knox:1995dq,Bull:2014rha,Durrer:2020orn,Jolicoeur:2020eup} 
\begin{align} \label{eqn:Nhi_dish}
    \sigma_{\rm HI}(\nu_i) &= \frac{4\pi f_{\rm sky}\,T^2_{\rm sys}(\nu_i)}{2N_{\rm dish}\, t_{\rm tot}\, \Delta\nu} \,,\\
    T_{\rm sys}(\nu_i) &= 25 + 60\left(\frac{300\,\text{MHz}}{\nu_i}\right)^{2.55}~ \text{K} \,,
\end{align}
where $t_{\rm tot}$ is the total observing time.
We assume the noise is deconvolved with a Gaussian beam, modelled as
\begin{equation}
    {{\cal N}_\ell^{\rm HI}}(\nu_i) = \sigma_{\rm HI}(\nu_i)\, B_\ell^{-2}(\nu_i) \,,
\end{equation}
with
\begin{equation} \label{eqn:beam}
    B_\ell = \exp\left[-\ell(\ell+1)\frac{\theta^2_{\rm FWHM}}{16 \ln 2}\right] \,,
\end{equation}
and
\be
\theta_{\rm FWHM} = \frac{1.22\, \lambda_i}{D_{\rm dish}}\quad\mbox{where}\quad\lambda_i=\lambda_{21}(1+z_i) \,.
\ee

For the next-generation SKAO-MID, we follow the SKAO Cosmology Red Book \citep{Bacon:2018dui} and use $N_{\rm dish} = 197$, $D_{\rm dish} = 15$\,m, $t_{\rm tot} = 10^4$\,hr, observing over 20,000\,deg$^2$ in the redshift range $0.35 \le z \le 3.05$ 
($1050 \geq \nu \geq 350\,$MHz, Band 1). We divide  the redshift range into 27 tomographic bins 
with  width 0.1. The cleaning of foregrounds from the HI intensity map effectively removes the largest scales,  $\ell \lesssim 5$ \citep{Witzemann:2018cdx,Cunnington:2019lvb}. and we take $\ell_{\rm min}=5$.

\subsection{Radio survey: interferometer mode}
For interferometer-mode intensity mapping, the noise is \citep{Bull:2014rha,Alonso:2017dgh,Ansari:2018ury,Durrer:2020orn,Jolicoeur:2020eup}
\begin{align} \label{eqn:Nhi_int}
    {{\cal N}_\ell^{\rm HI}}(\nu_i) &= \frac{4\pi f_{\rm sky}\,T^2_{\rm sys}(\nu_i)}{2N_{\rm dish}\, t_{\rm tot}\, \Delta\nu}\, \frac{\theta_{\rm FWHM}^2(\lambda_i)}{\eta^2\, N_{\rm b}\left(\ell \lambda_i/(2\pi)\right)\lambda_i^2} \,,\\
    T_{\rm sys}(\nu_i) &= \widetilde{T}_{\rm ampl} + \widetilde{T}_{\rm ground} + T_{\rm sky} \,,
\end{align}
where $\eta=0.7$ is the aperture efficiency factor, $N_{\rm b}$ is the density  of baselines in the image plane,
$\widetilde{T}_{\rm ampl} = 61.73$ K 
is the amplifier noise temperature corrected by the optical efficiency, 
$\widetilde{T}_{\rm ground} = 33.33$ K is due to the fraction of primary beam 
hitting the ground, and
\begin{equation}
    T_{\rm sky}(\nu_i) = 2.7 + 25\left(\frac{400\,\text{MHz}}{\nu_i}\right)^{2.75}~ \text{K} \,.
\end{equation}

For the futuristic PUMA experiment, we follow \citet{Ansari:2018ury,PUMA:2019jwd} and assume $N_{\rm dish} = 32,000$ 
dishes arranged in hexagonal close-packed array with 50\% fill factor, $D_{\rm dish} = 6$\,m,
an integration time of $t_{\rm tot} = 4\times10^4$\,hr, 
observing over half of the sky ($f_{\rm sky}=0.5$) 
in the redshift range $0.3 \le z \le 6$.
We divide  the redshift range into 57 tomographic bins with  width 0.1. 
As for SKAO, we take $\ell_{\rm min}=5$.

\subsection{Optical survey}
For a next-generation photometric galaxy survey similar to the Vera C. Rubin
Observatory’s Legacy Survey of Space and Time (LSST), we assume 
a redshift distribution of sources of the form
\begin{equation} \label{eqn:dNdz}
    \bar{n}_g(z) \propto z^{\alpha} \exp \left[ -\left( \frac{z}{z_0} \right)^{\beta} \right] ~\mbox{gal/arcmin}^2\,.
\end{equation}
The distribution of sources in the $i$-th redshift bin, including 
photometric uncertainties, following \cite{Ma:2005rc}, is
\begin{equation}
    n^i_g(z) = \int_{z^i_{\rm ph}}^{z^{i+1}_{\rm ph}} \dd z_{\rm ph}\, \bar{n}_g(z)\, p(z_{\rm ph}|z)\,,
\end{equation}
where we adopt a Gaussian distribution for the probability distribution of photometric redshift 
estimates $z_{\rm ph}$, given true redshifts $z$:
\begin{equation}
    p(z_{\rm ph}|z) = \frac{1}{\sqrt{2\pi}\,\sigma_z} \exp \Bigg[-\frac{\big(z-z_{\rm ph}\big)^2}{2\sigma_z^2}\Bigg] \,.
\end{equation}
The shot noise for galaxies in the $i$-th redshift bin is the inverse of the angular number density of galaxies: 
\begin{equation}
   {\cal N}^{gi}_\ell = \left( \int \dd z\ n^i_g(z)\right)^{-1}\,.
\end{equation}

For LSST clustering measurements, we assume a total number density of galaxies of $\bar{n}_g=48$ 
sources per arcmin$^2$, observed over 13,800\,deg$^2$ and distributed in redshift according to 
\eqref{eqn:dNdz}, with $\alpha$ = 2, $\beta$ = 0.9, and $z_{0}$ = 0.28, 
{corresponding to the Y10 gold sample ($i_{\rm lim} = 25.3$) specifications} from \cite{Mandelbaum:2018ouv}.
We assume 10 tomographic bins spaced by $0.1$, in the range $0.2 \leq z \leq 1.2$, with 
photometric redshift uncertainties $\sigma_z = 0.03 (1+z)$. The fiducial model for the bias is $b_g(z)= 0.95/D(z)$, 
where $D$ is the growth factor \citep{Mandelbaum:2018ouv}. We impose $\ell_{\rm min}=20$.

\subsection{Microwave survey}
We work with a possible CMB-S4 configuration assuming a 3 arcmin beam and 
$\sigma_{\rm T}^{1/2}=\sigma_{\rm P}^{1/2}/\sqrt{2}=1\,\mu$K-arcmin noise \citep{Abazajian:2016yjj}. 
We assume $\ell_{\rm min}=30$ and a different cut at high-$\ell$ of $\ell_{\rm max}^{\rm T}=3000$ 
in temperature and $\ell_{\rm max}^{\rm P}=5000$ in polarization, with $f_{\rm sky} = 0.4$.

For CMB temperature and polarization  angular power spectra, the instrumental noise 
deconvolved with the instrumental beam is defined by \citep{Knox:1995dq}
\begin{equation}
    {\cal N}_\ell^{\rm T,P} = \sigma_{\rm T,P}\, B_\ell^{-2}\,,
\end{equation}
where the Gaussian beam is given by \eqref{eqn:beam}.

For CMB lensing, we assume that the lensing reconstruction can be performed with the minimum 
variance quadratic estimator on the full sky, combining the TT, EE, BB, TE, TB, and EB  estimators, 
calculated according to \cite{Hu:2001kj} with {\tt quicklens}\footnote{\href{https://github.com/dhanson/quicklens}{https://github.com/dhanson/quicklens}} and 
applying iterative lensing reconstruction \citep{Hirata:2003ka,Smith:2010gu}.
We use the CMB-S4 lensing information in the range $30 \leq \ell \leq 3000$.

{We will refer to the full set of CMB information including temperature, 
E-mode polarization, CMB lensing, and their cross-correlations, as simply `CMB'.}

\section{Correlation between CMB lensing and large-scale structure} \label{sec:cross}
\begin{figure}
\centering\includegraphics[width=1\linewidth]{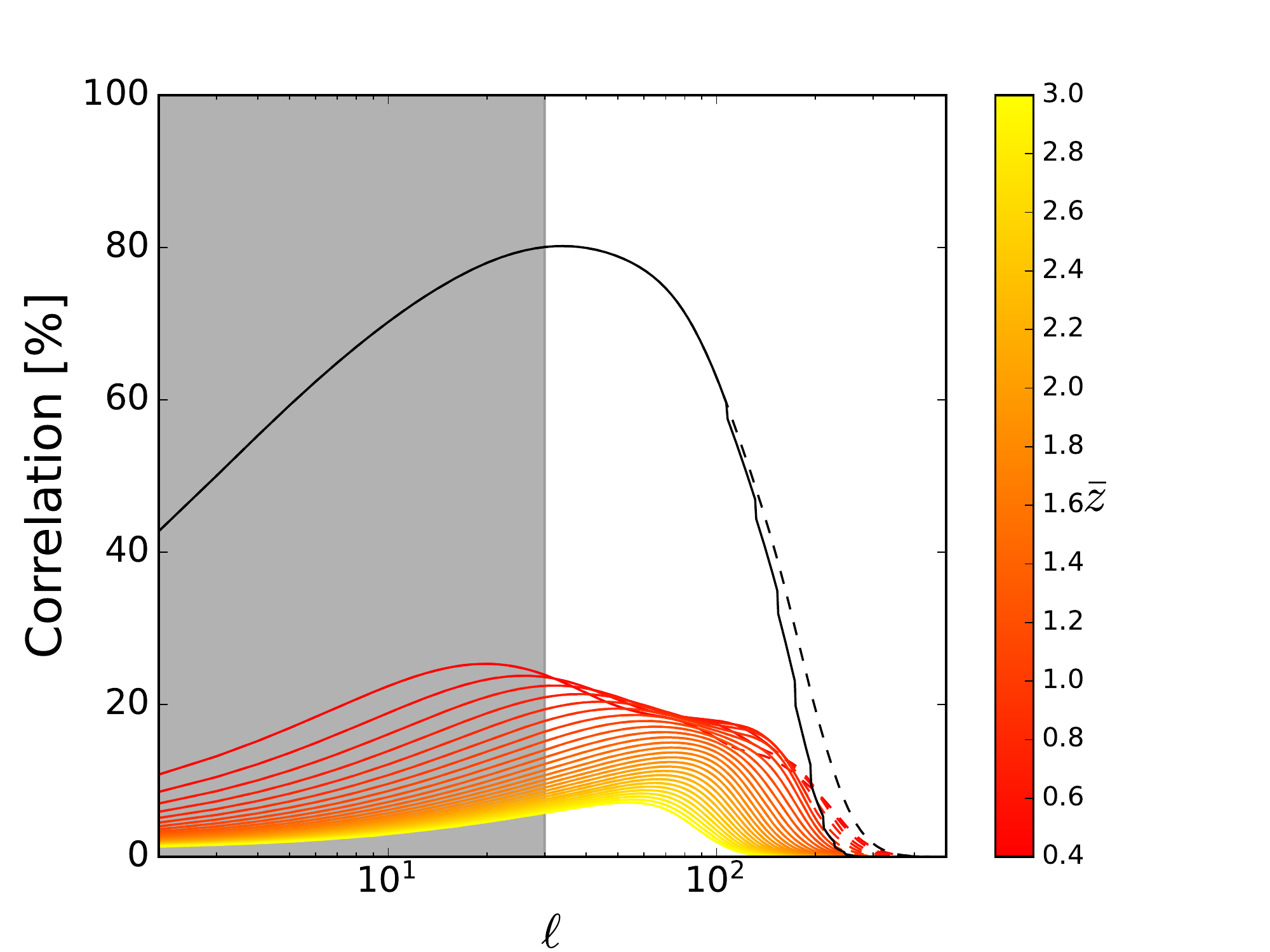}
\centering\includegraphics[width=1\linewidth]{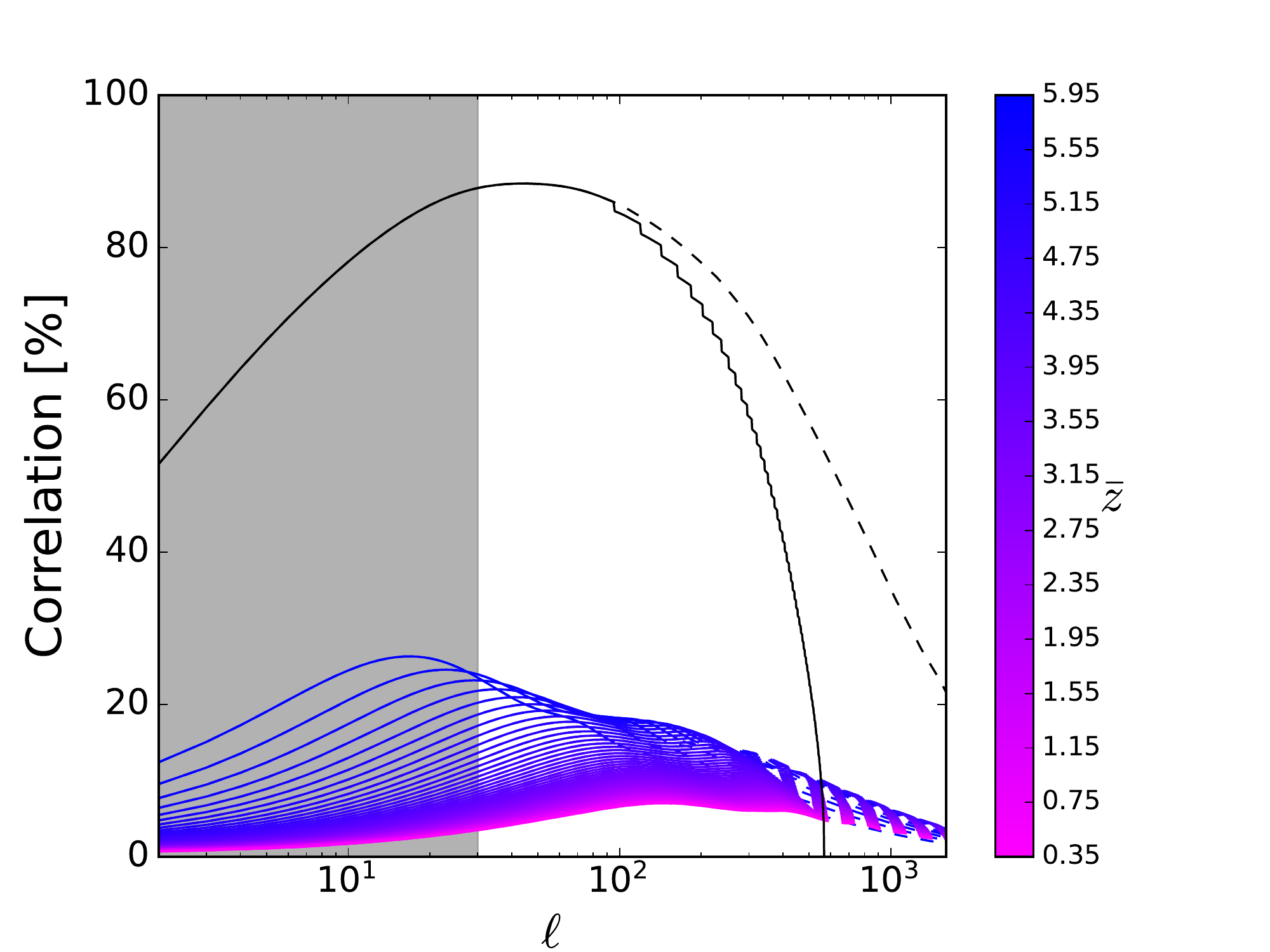}
\centering\includegraphics[width=1\linewidth]{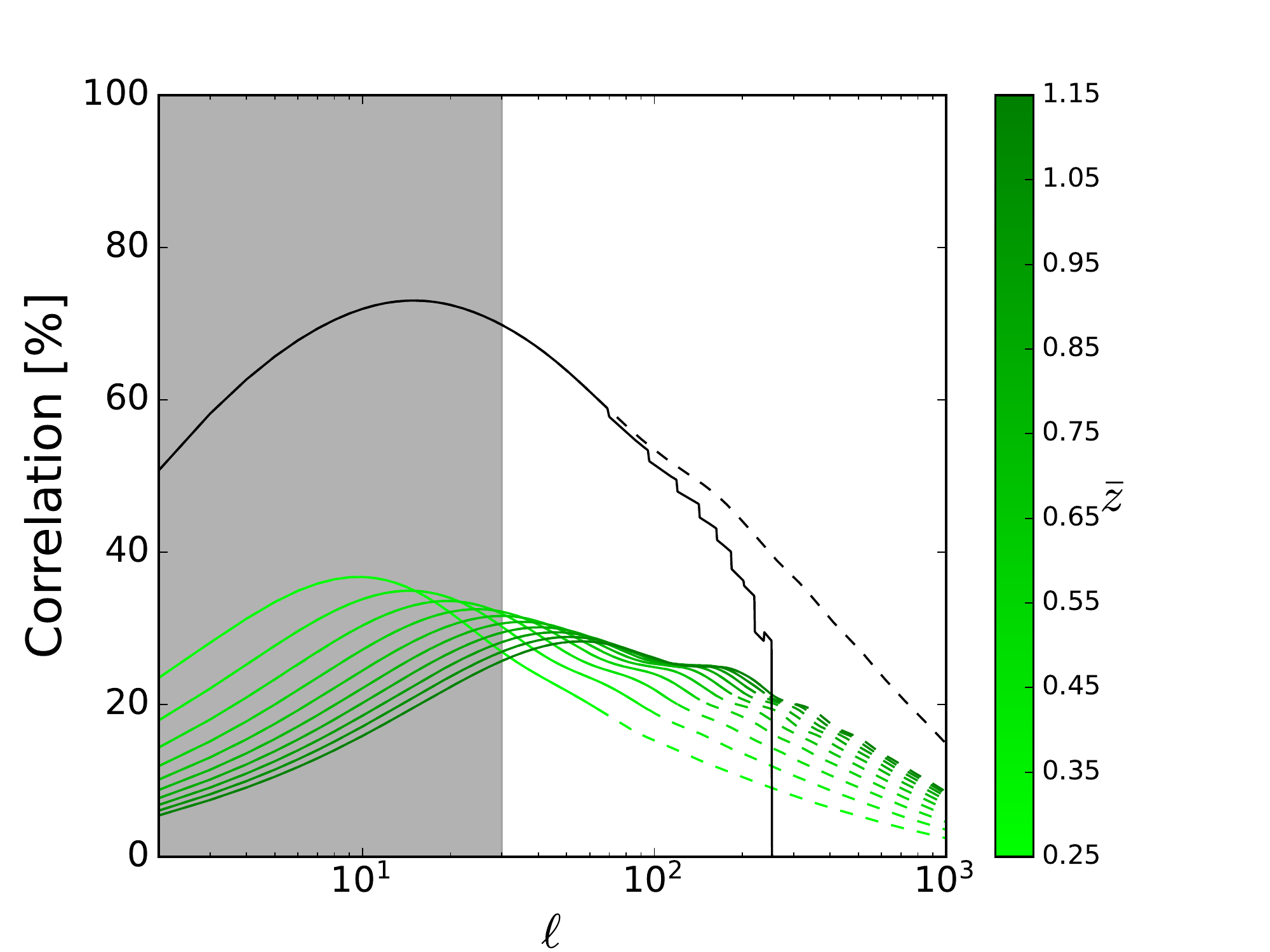}
\caption{Expected correlation coefficient \eqref{eqn:correlation} between CMB-S4 lensing 
and SKAO-MID redshift bins ({\em top}), PUMA redshift bins ({\em centre}), and LSST photometric redshift bins ({\em bottom}). Black line shows the combined coefficient.
{Dashed lines denote the correlation coefficient calculated 
without imposing the cut of non-linear scales corresponding to $\ell(z) \sim 0.1\,h\chi(z)$}. 
}
\label{fig:correlation}
\end{figure}

The performance of the cross-correlation analysis depends on the cross-correlation coefficient. 
This is similar to the signal-to-noise ratio in each redshift bin \citep{Ballardini:2018cho}, 
but without taking into account the number of modes and the sky fraction:
\begin{equation} \label{eqn:correlation}
    r^{X\phi}_\ell(z_i) = \frac{\left|C_\ell^{X\phi}(z_i)\right|}{\left[\Gamma_\ell^{XX}(z_i)\,\Gamma_\ell^{\phi\phi}\right]^{1/2}} \,,
\end{equation}
where $X=\,$ HI or g.
\autoref{fig:correlation} shows the correlation coefficient of SKAO-MID, PUMA, and LSST redshift bins 
with CMB lensing expected from CMB-S4. The grey shading for  $\ell<30$ is the region where we do not 
have cross-correlation with CMB-S4.

The cross-correlation reaches a maximum at low redshift and it drops on small scales. Moving toward 
higher redshift, the correlation peaks at higher multipoles. 
This is connected to the position of
the peak of the matter power spectrum in Fourier space at $k_{\rm peak} \sim 0.02\, h/$Mpc, which is 
mapped to higher multipoles for higher redshift according to $\ell_{\rm peak}(z) \simeq k_{\rm peak}\,\chi(z)$  where $\chi(z)$ is the comoving radial distance.

The correlation coefficient and in particular the one obtained combining the tomographic redshift bins 
(black line in \autoref{fig:correlation}) can be maximised by optimising the size and edges of 
the redshift bins.
Choosing these weights such that they maximise the correlation coefficient between the joint 
analyses can be used to maximise the effect of sample variance cancellation.

{Combining together the information from all redshift bins for each multipole, 
we can define the correlation coefficient \citep{Sherwin:2015baa}
\begin{equation}
r_\ell^{X\phi} = \left\{\sum_{i,j}\,  r_\ell^{X\phi}(z_i) \left[ r_\ell^{XX}\right]^{-1}\!\!(z_i,z_j)\,r_\ell^{X\phi}(z_j)\right\}^{1/2}
\end{equation}
where 
\begin{eqnarray}
 r_\ell^{XX}(z_i,z_j) = {C_\ell^{XX})(z_i,z_j) \over \Gamma_\ell^{XX}(z_i,z_j)}. 
\end{eqnarray}}
The 
{combined} cross-correlation coefficient for SKAO-MID and PUMA is higher 
($\sim 90\%$ at $\ell \sim 50$) in the region of interest compared to LSST. 
This is due to the wider redshift range probed by the  IM surveys compared to photometric surveys.
Note that the cross-correlation coefficient for SKAO-MID drops at $\ell \sim 200$ because of the telescope beam 
\eqref{eqn:beam}.

\section{Results} \label{sec:results}
We present in this section the uncertainties on the total neutrino mass $M_\nu$ 
for different combinations 
of cosmological surveys and considering a  conservative $k_{\rm max} = 0.1\, h/$Mpc, so that we consider on scales where linear perturbation theory is reliable. 
(Note that for SKAO-MID,  the telescope beam effectively removes scales $k>0.1\, h/$Mpc).
This $k$-cut is propagated to angular modes through the relation 
$\ell(z) \sim k\chi(z)$. 

Note that CMB-S4 with $\ell_{\rm min} = 30$ delivers $\sigma\left(M_\nu \right) \simeq 115$\,meV 
while complementing the low-$\ell$ down to $\ell_{\rm min} = 
{2}$ with LiteBIRD, it 
can reach $\sigma\left(M_\nu \right) \simeq 38\,$meV.

Uncertainties are marginalised over all  6 standard 
cosmological parameters in \eqref{parvec}. 
We also marginalise over the nuisance
parameters that allow for a free redshift evolution of the clustering bias $b_g$, or of the combination 
$T_{\rm HI}b_{\rm HI}$ for IM, and the magnification bias $s_g$, for each redshift bin. This leads to 27 
temperature-bias parameters for SKAO-MID, 57 temperature-bias parameters for PUMA,  10 clustering bias parameters  
and 10 magnification bias parameters for the LSST survey.

\begin{table}
\centering
\begin{tabular}{l|cc}
\hline
\hline
 & \multicolumn{2}{|c|}{$\sigma(M_\nu)$ [meV]} \\
 & without bias  & with bias \\
\hline
SKAO-MID                   & 266     & 234 \\
PUMA                       & 76      & 77  \\
LSST                       & 846     & 782 \\
\hline
SKAO-MID $\times$ CMB-S4   & 47      & 45 \\
PUMA $\times$ CMB-S4       & 30      & 26 \\
LSST $\times$ CMB-S4       & 69      & 62 \\
\hline
SKAO-MID $\times$ LSST     & 240     & 118 \\
PUMA $\times$ LSST         & 59      & 54  \\
\hline
\hline
\end{tabular}
\caption{
Two-tracer case. Marginalised uncertainties on $M_\nu$ at 68\% CL including (right column) or neglecting 
(left column) the scale-dependent halo bias induced by massive neutrinos. Uncertainties are for $k_{\rm max} = 0.1\, h/$Mpc.}
\label{tab:err}
\end{table}

The uncertainties for the single surveys with SKAO-MID, PUMA and LSST are large compared to the 
fiducial assumption of $M_\nu = 60$ meV, as shown in \autoref{tab:err}. Single-tracer results do 
improve when the scale-dependent bias is included, but the major improvement comes from the multi-tracer. 
Including CMB information from CMB-S4 with $\ell_{\rm min}=30$, using the multi-tracer, delivers: 
\begin{equation}
\sigma\left( M_\nu \right) \simeq \begin{cases} 45\ {\rm meV} & \mbox{SKAO-MID}\times \mbox{CMB-S4}\,, \\ 
26\ {\rm meV} & \mbox{PUMA}\times \mbox{CMB-S4} \,, \\
62\ {\rm meV} & \mbox{LSST}\times \mbox{CMB-S4} \,,
\end{cases}
\end{equation}
for $k_{\rm max} = 0.1\, h/$Mpc, 
{while the combination of IM and LSST leads to
\begin{equation}
\sigma\left( M_\nu \right) \simeq \begin{cases} 118\ {\rm meV} & \mbox{SKAO-MID}\times \mbox{LSST} \,, \\
54\ {\rm meV} & \mbox{PUMA}\times \mbox{LSST} \,.
\end{cases}
\end{equation}
}

When all three tracers are combined, the tightest constraints obtained are
\begin{equation}
\sigma\left( M_\nu \right) \simeq \begin{cases} 
12\ {\rm meV} & \mbox{SKAO-MID}\times\mbox{LSST}\times \mbox{CMB-S4}\,, \\
11 \ {\rm meV} & \mbox{PUMA}\times\mbox{LSST}\times \mbox{CMB-S4} \,, 
\end{cases}
\end{equation}
for $k_{\rm max} = 0.1\, h/$Mpc.

\begin{figure}
\includegraphics[width=.45\textwidth]{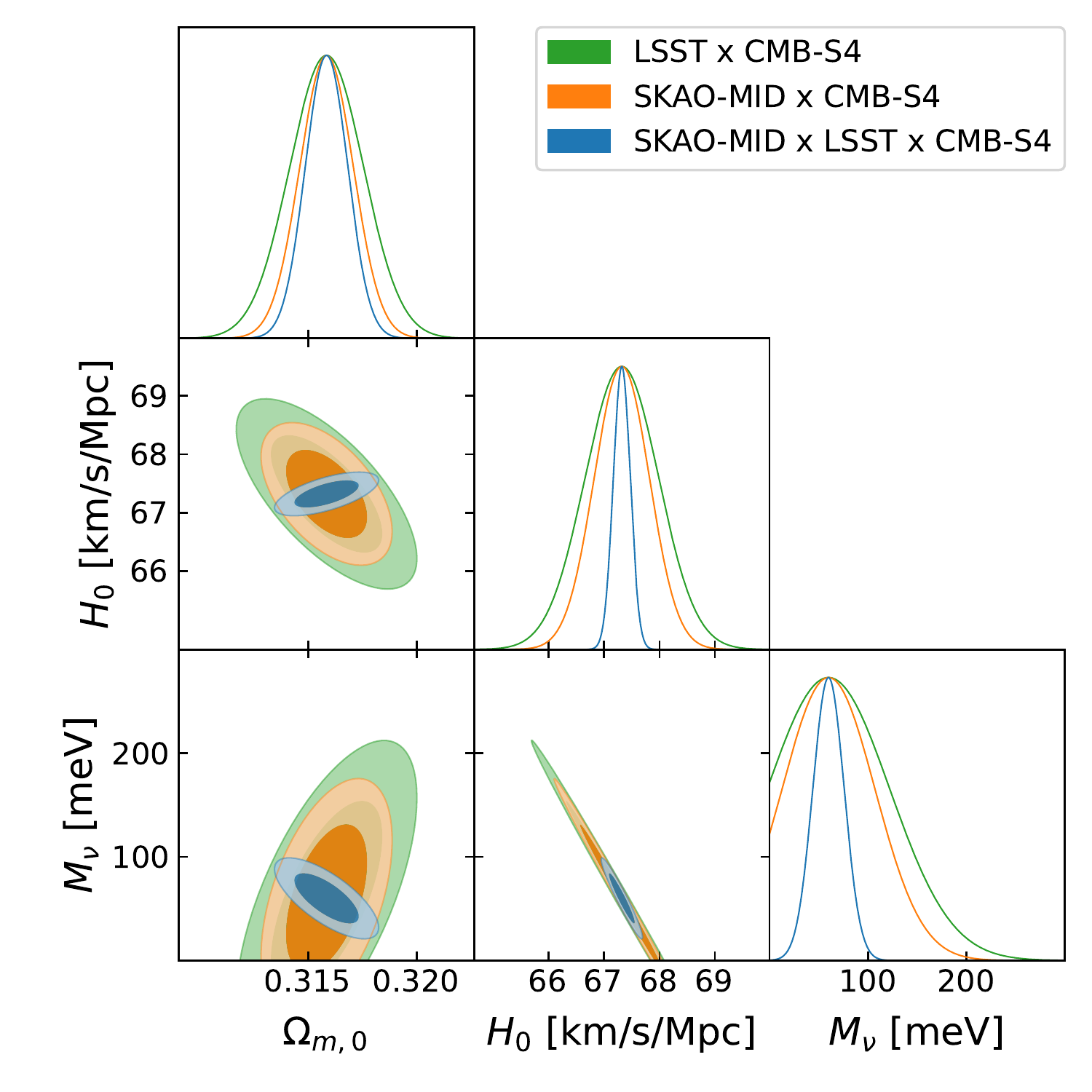}
\caption{
{Marginalised 2-dimensional contours (68\% and 95\% CL) for 
$\Omega_{m,0}$, $H_0$, $M_\nu$.
The multi-tracer combinations are: LSST x CMB-S4 (green), SKAO-MID × CMB-S4 (orange), 
and SKAO × LSST x CMB-S4 (blue).}}
\label{fig:triangle}
\end{figure}

We present  the three-tracer case in \autoref{tab:err_multi}. 
Adding CMB-S4 information to the multi-tracer combination of SKAO-MID and LSST reduces the 
error by a factor $\sim2$, to 63\,meV. 
However, if we include all cross-correlations -- between the CMB fields, intensity mapping 
and number counts -- then the error is reduced by a factor $\sim5$, to 12\,meV.
This effect is completely due to the cross-correlation with the CMB lensing $\phi$: 
neglecting the cross-correlation with E and T, we found the change in the uncertainties on 
the neutrino mass to be less than 1\% in all cases. 
{Figure~\ref{fig:triangle} shows the marginalized uncertainties on the 
3-dimensional $(\Omega_{m,0},\, H_0,\, M_\nu)$ parameter space.}

\begin{table}
\centering
\begin{tabular}{l|cc}
\hline
\hline
& \multicolumn{2}{|c|}{$\sigma(M_\nu)$ [meV]} \\
 & without bias  & with bias \\
\hline
SKAO-MID $\times$ LSST + CMB-S4         & 73  & 63 \\
SKAO-MID $\times$ LSST $\times$ CMB-S4  & 17  & 12 \\
PUMA $\times$ LSST + CMB-S4             & 28  & 28 \\
PUMA $\times$ LSST $\times$ CMB-S4      & 12  & 11 \\
\hline
\hline
\end{tabular}
\caption{
As in \autoref{tab:err}, for the three-tracer case.}
\label{tab:err_multi}
\end{table}

As already shown in \cite{Brinckmann:2018owf,Yu:2018tem}, the addition of extra constraints 
on the optical depth $\tau_{\rm reio}$ at recombination can  reduce further the uncertainties on the 
neutrino mass, breaking the partial degeneracy between $M_\nu$ and the normalization of the 
anisotropy $A_s \exp(-2\tau_{\rm reio})$. We present in Table~\ref{tab:err_multi_tau} the uncertainties 
on $M_\nu$ obtained when adding a Gaussian prior of $\sigma(\tau_{\rm reio}) = 0.008$, corresponding to 
current constraints from $Planck$ \citep{Aghanim:2018eyx}. We  also consider 
$\sigma(\tau_{\rm reio}) = 0.001$, which is possible with future CMB cosmic-variance polarisation 
experiments, or using independent information from 21cm IM \citep{Liu:2015txa}.
\begin{table}
\centering
\begin{tabular}{l|cc}
\hline
\hline
 & \multicolumn{2}{|c|}{$\sigma(M_\nu)$ [meV]} \\
\hline
$\sigma(\tau_{\rm reio})$                &  0.008 & 0.001 \\
\hline 
SKAO-MID $\times$ LSST $\times$ CMB-S4   & 14  & 11 \\
PUMA $\times$ LSST $\times$ CMB-S4       & 11  & 9 \\
\hline
\hline
\end{tabular}
\caption{
As in \autoref{tab:err_multi}, including the scale-dependent halo bias 
in combination with a Gaussian prior on $\tau_{\rm reio}$.}
\label{tab:err_multi_tau}
\end{table}

Note that, considering our agnostic assumption about bias parameters any additional information 
regarding the clustering bias would improve the constraints on the neutrino mass.
By contrast, marginalisation over magnification bias  does not significantly affect uncertainties 
on neutrino mass.

Finally, we find that modelling the scale-dependent features in the clustering bias due to 
the presence of massive neutrinos reduces the uncertainties significantly for the single-tracer cases, 
but  for the multi-tracer cases there is only a slight improvement for such a small choice of $M_\nu = 60$ meV.

\section{Conclusions} \label{sec:conc}
{We presented Fisher forecast constraints on the total neutrino mass from upcoming 21cm intensity mapping and photometric galaxy surveys, together with CMB lensing, temperature and polarisation data.
We included the scale-dependent clustering bias that is induced by neutrinos. The critical feature of our analysis was to use a multi-tracer analysis that combined all the information, leading to significant improvements over single-tracer constraints or the simple  addition of separate Fisher information for single tracers. The best results are achieved with a 3-tracer combination of 21cm intensity mapping, photometric survey and CMB.}

Constraints on the neutrino mass have been extensively studied; see e.g. 
\cite{Hall:2012kg,Allison:2015qca,Villaescusa-Navarro:2015cca,Schmittfull:2017ffw,Brinckmann:2018owf,Yu:2018tem,Boyle:2018rva,Archidiacono:2020dvx,Xu:2020fyg,Tanidis:2020byi,Bermejo-Climent:2021jxf,Sailer:2021yzm,Bayer:2021kwg}. 
While the combination of CMB and large-scale structure measurements is 
very powerful in breaking the geometrical degeneracy at background level, 
cosmological constraints depend strongly on the CMB low-$\ell$ measurements. 
Uncertainties of $\sigma(M_\nu) \simeq 20-30$ meV, corresponding to a 2-3$\sigma$ 
detection for a minimum neutrino mass $M_\nu = 60$ meV, have been forecast when combining CMB anisotropies and future survey number counts, and 
including the {\em Planck} constraint $\sigma(\tau_{\rm reio}) \simeq 0.008$ on reionisation
\citep{Aghanim:2018eyx}.
These uncertainties are robust for extensions of the $\Lambda$CDM concordance model, 
such as cosmologies where also $N_{\rm eff}$ and the dark energy parameters are varied 
\citep{Brinckmann:2018owf,Boyle:2018rva}.
Moreover, the combination of different late-time probes such as BAO, galaxy clustering, 
galaxy weak lensing, and CMB weak lensing allows robust constraints without 
including any information regarding $\tau_{\rm reio}$ \citep{Brinckmann:2018owf,Boyle:2018rva,Xu:2020fyg}.

However, smaller uncertainties of the order of $\sigma(M_\nu) \simeq 10$ meV have been 
predicted only in combination with a tighter constraint on the optical depth at 
recombination $\sigma(\tau_{\rm reio}) \simeq 0.001-0.002$ \citep{Brinckmann:2018owf,Yu:2018tem}.

Using a three-tracer approach that combines clustering measurements from two next-generation large-scale 
structure surveys together with CMB-S4 measurements, we showed that if only linear scales are included, 
the 21cm intensity mapping surveys outperform the photometric LSST survey. 

The multi-tracer combination of CMB-S4 with intensity mapping delivers 
\begin{equation}
\sigma\left( M_\nu \right) \simeq \begin{cases} 45\ {\rm meV} & \mbox{SKAO-MID}\times \mbox{CMB-S4}\,, \\ 
{26}\ {\rm meV} & \mbox{PUMA}\times \mbox{CMB-S4} \,, 
\end{cases}
\end{equation}
for $k_{\rm max} = 0.1\, h/$Mpc, and when marginalising over 6 standard cosmological parameters 
and over all nuisance parameters (27 for SKAO-MID and 57 for PUMA). 
These results are not sensitive to increasing $\ell_{\rm min}$ up to 20 since the extra constraining 
power comes from the cross-correlation between CMB lensing with intensity mapping.

When all three tracers are included in a multi-tracer analysis, the tightest uncertainties 
predicted are
\begin{equation}
\sigma( M_\nu ) \simeq  12/11\ {\rm meV} ~~ \mbox{SKAO-MID/PUMA} \times 
\mbox{LSST} \times \mbox{CMB-S4}\,.
\end{equation}
Improved constraints in the linear regime can be obtained if we assume a prior on the reionisation 
optical depth (see \autoref{tab:err_multi_tau}). 

In \cite{Chen:2021vba} the 1-loop power spectrum (allowing an increase in $k_{\rm max}$) and  the 
tree-level bispectrum are used to significantly improve constraints from the cross-correlation of 
LSST and CMB-S4 \citep[see also][for galaxy bispectrum constraints on $M_\nu$]{Chudaykin:2019ock,Kamalinejad:2020izi,Hahn:2020lou}. 
Here we have avoided the complexities of going beyond linear modelling, relying instead on the 
multi-tracer combination of the first-order power spectra of two large-scale structure surveys together with CMB.

\section*{Acknowledgements}
MB acknowledges financial contribution from the contract ASI/ INAF for the Euclid mission n.2018-23-HH.0.
RM is supported by the South African Radio Astronomy Observatory (SARAO) and the National Research Foundation (Grant No. 75415) and by the UK STFC Consolidated Grant ST/S000550/1.

\section*{Data Availability}
The data underlying this article will be shared on reasonable request to the corresponding author.

\bibliographystyle{mnras}
\bibliography{Biblio}

\end{document}